\documentclass[prd,superscriptaddress,twocolumn,nofootinbib,showpacs,preprintnumbers,floatfix]{revtex4}

\usepackage{mathrsfs}
\usepackage{amsfonts,amssymb,amsmath}
\usepackage{graphicx,epsfig}
\usepackage{multirow}

\begin{document}

\title{Prospects for Discovery of Supersymmetric No-Scale $\cal{F}$-$SU(5)$\\at The Once and Future LHC}

\author{Tianjun Li}

\affiliation{State Key Laboratory of Theoretical Physics, Institute of Theoretical Physics,
Chinese Academy of Sciences, Beijing 100190, P. R. China }

\affiliation{George P. and Cynthia W. Mitchell Institute for Fundamental Physics and Astronomy,
Texas A$\&$M University, College Station, TX 77843, USA }

\author{James A. Maxin}

\affiliation{George P. and Cynthia W. Mitchell Institute for Fundamental Physics and Astronomy,
Texas A$\&$M University, College Station, TX 77843, USA }

\author{Dimitri V. Nanopoulos}

\affiliation{George P. and Cynthia W. Mitchell Institute for Fundamental Physics and Astronomy,
Texas A$\&$M University, College Station, TX 77843, USA }

\affiliation{Astroparticle Physics Group, Houston Advanced Research Center (HARC),
Mitchell Campus, Woodlands, TX 77381, USA}

\affiliation{Academy of Athens, Division of Natural Sciences,
28 Panepistimiou Avenue, Athens 10679, Greece }

\author{Joel W. Walker}

\affiliation{Department of Physics, Sam Houston State University,
Huntsville, TX 77341, USA }


\begin{abstract}

We present the reach of the Large Hadron Collider (LHC) into the parameter space 
of No-Scale $\cal{F}$-$SU(5)$, starting our analysis with the current operating energy
of $\sqrt{s} = 7$~TeV, and extending it on through the bright future of a 14 TeV beam.  No-Scale $\cal{F}$-$SU(5)$
is a model defined by the confluence of the ${\cal F}$-lipped $SU(5)$ Grand Unified Theory,
two pairs of hypothetical TeV scale vector-like supersymmetric multiplets with origins in ${\cal F}$-theory, 
and the dynamically established boundary conditions of No-Scale Supergravity. 
When searching for a five standard deviation signal, we find that the CMS experiment at the $\sqrt{s}$ = 7 TeV LHC began to penetrate the phenomenologically viable parameter space of this model at just under 1 ${\rm fb}^{-1}$ of integrated luminosity, and that
the majority of this space remains intact, subsequent to analyses of the first 1.1 ${\rm fb}^{-1}$ of CMS data. On the contrary, the ATLAS experiment had not reached the $\cal{F}$-$SU(5)$ parameter space in its first 1.34 ${\rm fb}^{-1}$ of luminosity.
Since the CMS and ATLAS detectors have now each amassed a milestone of 5 ${\rm fb}^{-1}$ of collected luminosity,
the current LHC is presently effectively probing No-Scale ${\cal F}$-$SU(5)$. Upon the crossing of the 
5 ${\rm fb}^{-1}$ threshold, the 7 TeV LHC will have achieved five standard deviation discoverability for a unified gaugino
mass of up to about 532 GeV, a light stop of 577 GeV, a gluino of 728 GeV, and heavy squarks 
of just over 1 TeV.  Extending the analysis to include a future LHC center-of-mass beam energy of
$\sqrt{s}$ = 14 TeV, the full model space of No-Scale $\cal{F}$-$SU(5)$ should be visible to CMS at about
30 ${\rm fb}^{-1}$ of integrated luminosity. We stress that the $\cal{F}$-$SU(5)$ discoverability thresholds discussed here are contingent upon retaining only those events with nine jets or more for the CMS experiment and seven jets or more for the ATLAS experiment.

\end{abstract}


\preprint{ACT-10-11, MIFPA-11-32}

\maketitle


\section{Introduction}

The exploration of physics beyond the Standard Model by the Large Hadron Collider (LHC) at CERN has been steady since 2010, gathering data from proton-proton collisions at a center-of-mass beam energy of ${\sqrt s}$ = 7 TeV. The detectors have reached an integrated luminosity of 5 ${\rm fb}^{-1}$ at the end of 2011, and conservatively, are expected to attain 20 ${\rm fb}^{-1}$ by the end of 2012. Subsequent to 2012, the LHC is anticipated to feature an increased beam energy of ${\sqrt s}$ = 10-14 TeV. The first significant milestone of 1 ${\rm fb}^{-1}$ was achieved in 2011, in the process generating a multitude of comprehensive experimental and phenomenological analyses by the CMS Collaboration~\cite{Khachatryan:2011tk,Chatrchyan:2011bz,Chatrchyan:2011bj,Collaboration:2011ida,Chatrchyan:2011ek,Collaboration:2011qs,PAS-SUS-11-003} and the ATLAS Collaboration~\cite{daCosta:2011hh,daCosta:2011qk,Aad:2011ks,Aad:2011xm,Collaboration:2011jz,Aad:2011qa} on the first few data collections. 

The leading candidate for an extension to the Standard Model is Supersymmetry (SUSY), a natural solution to the gauge hierarchy problem. Supersymmetric Grand Unified Theories (GUTs) with gravity mediated supersymmetry breaking, known as minimal Supergravity (mSUGRA) and the Constrained Minimal Supersymmetric Standard Model (CMSSM), have been thoroughly evaluated against the first 1 ${\rm fb}^{-1}$ of data. However, studies have shown that most of the experimentally viable parameter space of mSUGRA and the CMSSM has been excluded by the initial LHC constraints (for example, see Ref.~\cite{Strumia:2011}). The severe attenuation of the mSUGRA and CMSSM model space imposed by the early LHC results stimulates the question of whether SUSY and/or superstring post-Standard Model extensions exist that do unequivocally evade the LHC constraints thus far imposed, while remaining within the present and near-term reach of the LHC.

We investigate here such a model, that being No-Scale $\cal{F}$-$SU(5)$ with vector-like particles~\cite{Li:2010ws, Li:2010mi,Li:2010uu,Li:2011dw, Li:2011hr, Maxin:2011hy, Li:2011xu, Li:2011in,Li:2011gh,Li:2011fu,Li:2011xg,Li:2011ex,Li:2011av,Li:2011ab}.  The full No-Scale $\cal{F}$-$SU(5)$ model space allowed by a set of ``bare minimal'' experimental constraints was studied in~\cite{Li:2011xu}, which we apply as our baseline parameter space in this work. As we shall show, when searching for a five standard deviation signal, this model began to be probed by the CMS experiment at the LHC at just under 1 ${\rm fb}^{-1}$ of luminosity at ${\sqrt s}$ = 7 TeV. Thus, most of the $\cal{F}$-$SU(5)$ model space has remained wholly intact through accumulation of the first 1.1 ${\rm fb}^{-1}$ of data, an accomplishment certainly not shared by mSUGRA or the CMSSM. With recent news of each of the CMS and ATLAS experiments reaching 5 ${\rm fb}^{-1}$, the penetration of the LHC into the viable $\cal{F}$-$SU(5)$ parameter space has reached a modest level. Nonetheless, if a SUSY discovery is not imminent at ${\sqrt s}$ = 7 TeV, we shall show that coverage of the entire No-Scale $\cal{F}$-$SU(5)$ model space will only be achieved after the scheduled escalation to ${\sqrt s}$ = 14 TeV in 2013 or beyond.


\section{The  No-Scale $\cal{F}$-$SU(5)$ Model}


In the traditional framework, 
supersymmetry is broken in 
the hidden sector, and then its breaking effects are
 mediated to the observable sector
via gravity or gauge interactions. In the mSUGRA and CMSSM,
the supersymmetry breaking soft terms can be parameterized
by four universal parameters plus the sign of the Higgs bilinear 
mass term $\mu$: gaugino mass $M_{1/2}$,
 scalar mass $M_0$, trilinear soft term $A$, and
the ratio of Higgs vacuum expectation values (VEVs) 
$\tan\beta$ at low energy. The $\mu$ term and its bilinear 
 soft term $B_{\mu}$ are determined
by the $Z$-boson mass $M_Z$ and $\tan\beta$ after
the electroweak symmetry breaking. To solve the cosmological constant
problem, No-Scale supergravity is 
proposed~~\cite{Cremmer:1983bf,Ellis:1983sf, Ellis:1983ei, Ellis:1984bm, Lahanas:1986uc}. 
For the simplest stringy No-Scale supergravity, we obtain the 
boundary condition $M_0=A=B_{\mu}=0$~\cite{Ellis:1984bm, Lahanas:1986uc} while $M_{1/2}$ could be
non-zero at the unification scale. 

The No-Scale $\cal{F}$-$SU(5)$ models~\cite{Li:2010ws, Li:2010mi,Li:2010uu,Li:2011dw, Li:2011hr, Maxin:2011hy, Li:2011xu, Li:2011in,Li:2011gh,Li:2011fu,Li:2011xg,Li:2011ex,Li:2011av,Li:2011ab}
represent the unification of the ${\cal F}$-lipped $SU(5)$ Grand Unified Theory
(GUT)~\cite{Barr:1981qv,Derendinger:1983aj,Antoniadis:1987dx},
two pairs of hypothetical TeV scale vector-like supersymmetric multiplets with origins in
${\cal F}$-theory~\cite{Jiang:2006hf,Jiang:2009zza,Jiang:2009za,Li:2010dp,Li:2010rz},
and the dynamically established boundary conditions of No-Scale
Supergravity~\cite{Cremmer:1983bf,Ellis:1983sf, Ellis:1983ei, Ellis:1984bm, Lahanas:1986uc}.
For a thorough review on  the No-Scale
$\cal{F}$-$SU(5)$ model,
the reader is directed to the appendix of Ref.~\cite{Maxin:2011hy}.

Employing No-Scale boundary conditions at the $\cal{F}$-$SU(5)$ unification
scale, we have illustrated an exceptionally constrained ``golden point''~\cite{Li:2010ws} 
and ``golden strip''~\cite{Li:2010mi,Li:2011xg} which satisfy the most recent 
experimental constraints, while featuring also an imminently observable proton 
decay rate~\cite{Li:2009fq}.  The boundary constraint on $B_{\mu}=0$ is, in particular,
very severe, drastically condensing the viable parameter space.
Furthermore, by utilizing a Super No-Scale
condition, we have dynamically determined $M_{1/2}$ and $\tan\beta$~\cite{Li:2010uu,Li:2011dw,Li:2011xu,Li:2011ex}.
The modulus was thus stabilized dynamically~\cite{Li:2010uu, Li:2011dw,Li:2011xu,Li:2011ex} due to the fact that $M_{1/2}$ is related to the modulus field of the internal space in string models.

In the simplest No-Scale supergravity, 
all the supersymmetry breaking soft terms arise from a single 
parameter $M_{1/2}$, therefore, the particle spectra are proportionally similar
up to an overall rescaling on $M_{1/2}$, hence leaving invariant most of the ``internal'' 
physical properties.  This rescaling ability on $M_{1/2}$ is not apparent in alternative
supersymmetric models, due to their larger parameterization freedom, including in particular
that of $M_0$.  The rescaling symmetry can likewise be broken to a certain degree 
by the vector-like particle
mass parameter, although this effect is weak.

\begin{figure*}[htp]
	\centering
	\includegraphics[width=0.80\textwidth]{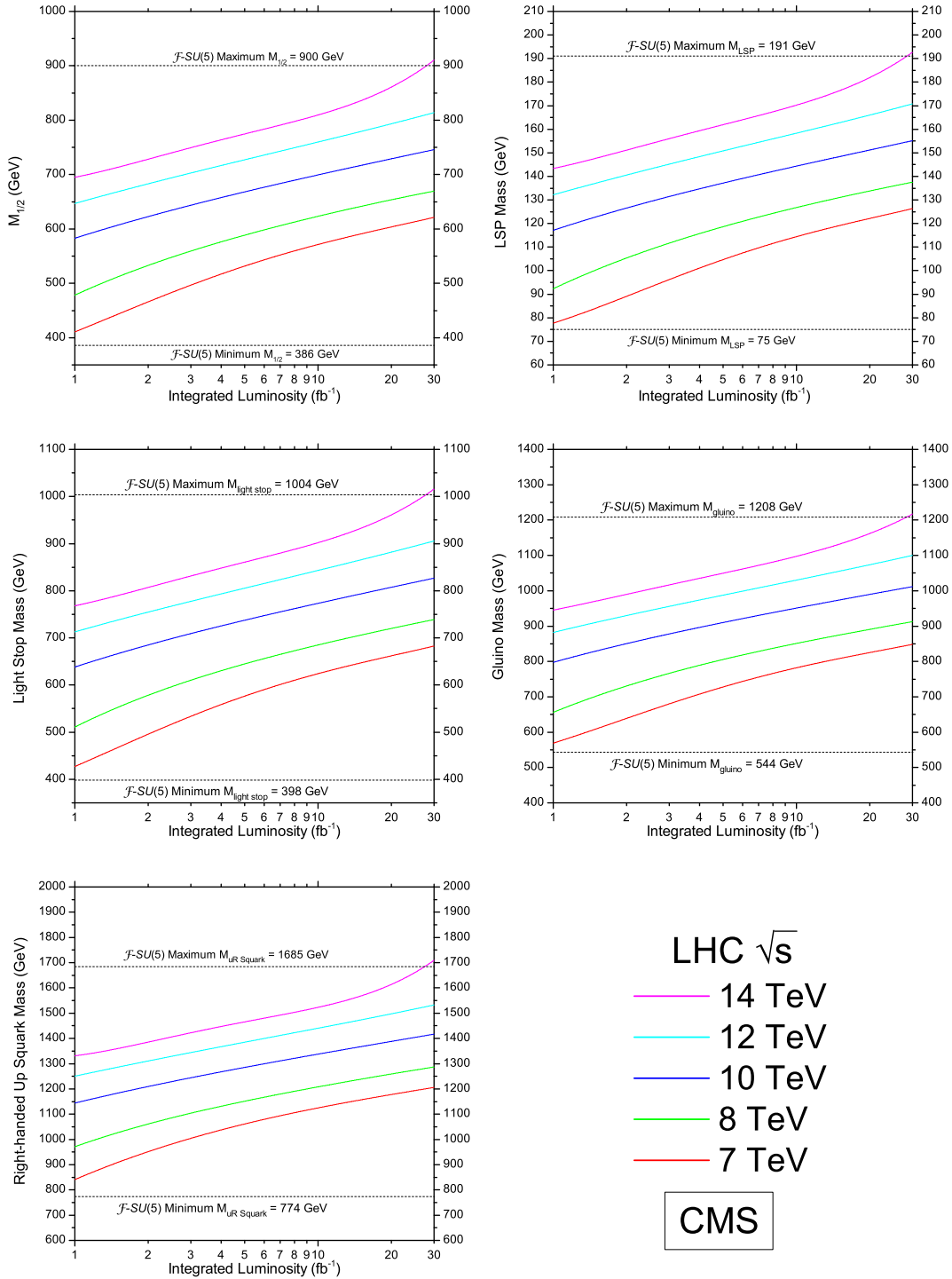}
	\caption{No-Scale $\cal{F}$-$SU(5)$ sparticle mass discovery threshold as a function of integrated luminosity utilizing the $S/\sqrt{B+1}\,\ge 5$ condition, using the CMS experiment post-processing cuts and Standard Model background sample of Reference~\cite{PAS-SUS-11-003}. We augment the CMS jet cutting strategy of~\cite{PAS-SUS-11-003} by retaining only those events with greater than or equal to nine jets. Annotated are the maximum and minimum SUSY masses allowable by the application of the ``bare minimal'' experimental constraints of~\cite{Li:2011xu}. Each plot space visibly depicts that the 1 ${\rm fb}^{-1}$ mass threshold is close to the minimum mass permitted within the No-Scale $\cal{F}$-$SU(5)$ model space, thus leaving the parameter space mostly intact after the first 1.1 ${\rm fb}^{-1}$ of LHC data. As shown, the No-Scale $\cal{F}$-$SU(5)$ began to be probed at just under 1 ${\rm fb}^{-1}$, however, the entire model space is not within reach of the ${\sqrt s}$ = 7 TeV LHC (bottom curve in each plot), with full coverage requiring a minimum of ${\sqrt s}$ = 14 TeV (top curve in each plot) for 30 ${\rm fb}^{-1}$, if a signal discovery is not accomplished at ${\sqrt s}$ = 7 TeV.}
	\label{fig:mass_luminosity_cms}
\end{figure*}

\begin{figure*}[htp]
	\centering
	\includegraphics[width=0.80\textwidth]{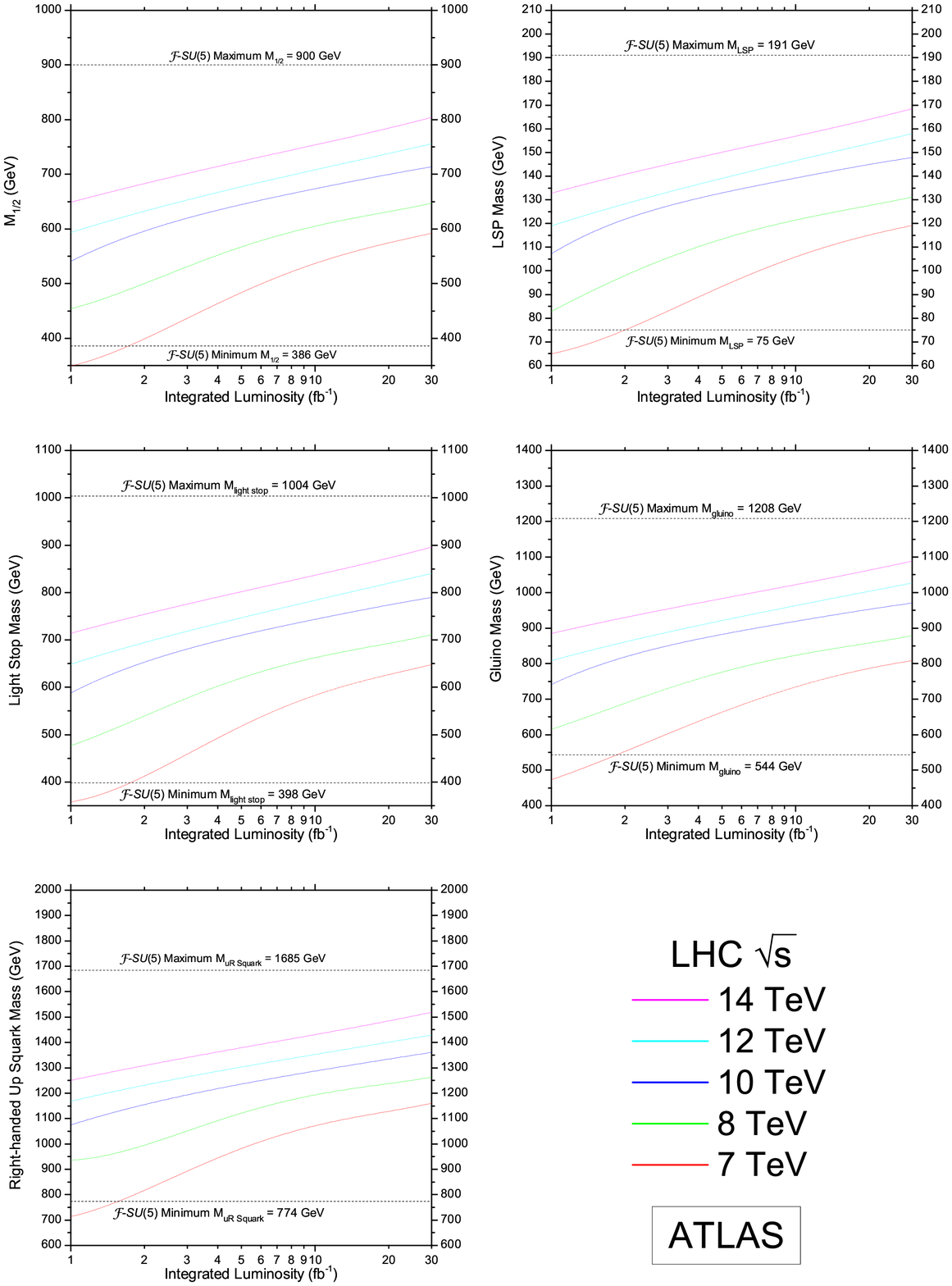}
	\caption{No-Scale $\cal{F}$-$SU(5)$ sparticle mass discovery threshold as a function of integrated luminosity utilizing the $S/\sqrt{B+1}\,\ge 5$ condition, using the ATLAS experiment post-processing cuts and Standard Model background sample of Reference~\cite{Aad:2011qa}. We augment the ATLAS jet cutting strategy of~\cite{Aad:2011qa} by retaining only those events with greater than or equal to seven jets. Annotated are the maximum and minimum SUSY masses allowable by the application of the ``bare minimal'' experimental constraints of~\cite{Li:2011xu}. Each plot space visibly depicts that the 1 ${\rm fb}^{-1}$ mass threshold is just below the minimum mass permitted within the No-Scale $\cal{F}$-$SU(5)$ model space, thus leaving the parameter space entirely intact after the first 1.34 ${\rm fb}^{-1}$ of LHC data. As shown, the No-Scale $\cal{F}$-$SU(5)$ will not begin to be probed until just under 2 ${\rm fb}^{-1}$, however, the entire model space is not within reach of the ${\sqrt s}$ = 7 TeV LHC (bottom curve in each plot), with full coverage requiring a minimum of ${\sqrt s}$ = 14 TeV (top curve in each plot) and much more than 30 ${\rm fb}^{-1}$, if a signal discovery is not accomplished at ${\sqrt s}$ = 7 TeV.}
	\label{fig:mass_luminosity_atlas}
\end{figure*}


\section{Supersymmetry Discovery}


We presented a SUSY cold dark matter Discovery Index in Reference~\cite{Li:2011gh}, drawing on a well known indicator of the statistical significance of a collider signal in relation to a competing background. The ratio $S/\sqrt{B+1}$ of signal events $S$ to the square root of background events $B$, plus one, was applied to the number of events for a No-Scale
 $\cal{F}$-$SU(5)$ signal with greater than or equal to nine jets~\cite{Li:2011hr,Maxin:2011hy,Li:2011fu}, as a function of the Lightest Supersymmetric Particle (LSP) mass.  In the event of very small background samples, the ``plus one''
limiter in the denominator prevents the signal count $S$ from being numerically smaller than the intended ratio.  The minimally favorable discovery ratio is generally considered to be in the vicinity of $S/\sqrt{B+1}\,\ge 5$, with much larger ratios more favorable.

We extend the Discovery Index study of~\cite{Li:2011gh} using the ratio $S/\sqrt{B+1}$ to evaluate not just the LSP, but also the unified gaugino mass $M_{1/2}$, light stop mass $m_{\widetilde{t}_1}$, gluino mass $m_{\widetilde{g}}$, and as an example of a heavy squark, the right-handed up squark mass $m_{\widetilde{u}_R}$. We strategically choose to focus on these sparticles due to the distinctive No-Scale $\cal{F}$-$SU(5)$ SUSY mass spectrum hierarchy of $m_{\tilde{t_1}} < m_{\tilde{g}} < m_{\tilde{q}}$, a possibly unique mass pattern, which is unseen in any of the ``Snowmass Points and Slopes'' (SPS) benchmark points~\cite{Allanach:2002nj}. The Snowmass benchmarks were specifically chosen by a consensus of proposals to represent a broad range of the most popular SUSY breaking mechanisms, and the exclusion of the aforementioned mass pattern is therefore highly indicative of its potential rarity. The logic of our derivation is that if we take the minimum ratio of five to signify a discovery threshold, then we can obtain from this ratio the minimum sparticle mass that is observable for a given integrated luminosity and center-of-mass beam energy, providing a detailed illustration of the reach of the LHC into the $\cal{F}$-$SU(5)$ parameter space.

Our Monte Carlo detector simulation has been performed using the {\tt MadGraph}~\cite{Stelzer:1994ta,MGME} suite, including
the standard {\tt MadEvent}~\cite{Alwall:2007st}, {\tt PYTHIA}~\cite{Sjostrand:2006za} and {\tt PGS4}~\cite{PGS4} chain, with 
post-processing performed by a custom script {\tt CutLHCO}~\cite{cutlhco} which implements the desired cuts,
and counts and compiles the associated net statistics.  All 2-body SUSY processes are included in our simulation.
We employ a proprietary modification of the {\tt SuSpect 2.34}~\cite{Djouadi:2002ze}
codebase to run the RGEs, and our SUSY particle spectrum computations are executed with {\tt MicrOMEGAs 2.1}~\cite{Belanger:2008sj}. Our Standard Model sample is borrowed from the estimated backgrounds derived by the CMS experiment~\cite{PAS-SUS-11-003} and the ATLAS experiment~\cite{Aad:2011qa}, which include all QCD, W-boson, Z-boson, and $t \overline{t}$ processes. For collision energies above the current $\sqrt{s} = 7$~TeV operational phase, we proportionally index the official collaboration backgrounds against the Monte Carlo scaling of $\cal{F}$-$SU(5)$. For the CMS analysis, we mimic the CMS post-processing cuts of Ref.~\cite{PAS-SUS-11-003} processed without the cut on the $\alpha_{\rm T}$ statistic and binned according to jet count, including the ultra-high jet multiplicities we have vigorously advocated~\cite{Li:2011hr,Maxin:2011hy,Li:2011fu,Li:2011av,Li:2011ab}, where we retain only those events with greater than or equal to nine jets. We have argued that the CMS $\alpha_{\rm T}$ cut is actually actively biased against high jet multiplicities, and that it is moreover unnecessary in this regime, given a substantial natural suppression of the background based simply on the jet count threshold itself. For the ATLAS analysis, in the same manner as the CMS analysis, we likewise mimic the ATLAS post-processing cuts of Ref.~\cite{Aad:2011qa} for the case of jet $p_T >$ 80 GeV and $H_{\rm T}^{\rm miss}/\sqrt{H_{\rm T}}$ in the range $(3.5 \to \inf)$. However, in the case of these ATLAS cuts, we retain only those events with greater than or equal to seven jets, due to the more suppressive nature of the ATLAS cuts in the high jet multiplicity regime. Nonetheless, even with the lower cut on the number of jets at seven for the ATLAS data, the $\cal{F}$-$SU(5)$ signal is discoverable at five standard deviations, though at a larger necessary integrated luminosity than the less suppressive CMS jet cutting strategy.

As exhibited in Figure Sets (\ref{fig:mass_luminosity_cms}-\ref{fig:mass_luminosity_atlas}), the logarithm of the integrated luminosity varies nearly linearly with the sparticle mass discovery threshold. We demarcate a range of luminosities from 1 ${\rm fb}^{-1}$ to 30 ${\rm fb}^{-1}$ in order to encompass the full operational range of the ``once and future'' LHC. Also annotated on each plot space is the maximum and minimum mass allowed using only the ``bare minimal'' experimental constraints of~\cite{Li:2011xu}. Each plot clearly shows that the SUSY mass discovery threshold for 1 ${\rm fb}^{-1}$ is near or comfortably below the minimum mass allowed within the No-Scale $\cal{F}$-$SU(5)$ model space. As a consequence, where the traditional SUSY models such as mSUGRA and the CMSSM have already experienced severe reductions in their experimentally viable parameter space, the No-Scale $\cal{F}$-$SU(5)$ has just begun to be penetrated. However, recent reports place the presently collected raw luminosities at 5 ${\rm fb}^{-1}$ for each of the CMS and ATLAS experiments.  Therefore, as Figure Sets (\ref{fig:mass_luminosity_cms}-\ref{fig:mass_luminosity_atlas}) depict, the No-Scale $\cal{F}$-$SU(5)$ parameter space is now currently being probed at the ${\sqrt s}$ = 7 TeV LHC.  However, only roughly half of the viable mass range can be probed by the ${\sqrt s}$ = 7 TeV LHC, and if no discovery occurs by the conclusion of 2012, we shall eagerly anticipate the planned commencement of the ${\sqrt s}$ = 14 TeV LHC era.  This upgraded machine would be capable of covering the entirety of the No-Scale $\cal{F}$-$SU(5)$ parameter space with 30 ${\rm fb}^{-1}$ of integrated luminosity. 

\begin{table}[ht]
	\centering
	\caption{Reach of the ${\sqrt s}$ = 7 TeV LHC into the No-Scale $\cal{F}$-$SU(5)$ model space utilizing the $S/\sqrt{B+1}\,\ge 5$ condition, using the CMS experiment post-processing cuts and Standard Model background sample of Reference~\cite{PAS-SUS-11-003}. We augment the CMS jet cutting strategy of~\cite{PAS-SUS-11-003} by retaining only those events with greater than or equal to nine jets. All masses are in GeV.}
		\begin{tabular}{|c|c|c|c|c|c|} \hline
${\rm Luminosity}$&$~~M_{1/2}~~$&$~~m_{\widetilde{\chi}^{0}_{1}}~~$&$~~m_{\widetilde{t}_{1}}~~$&$~~m_{\widetilde{g}}~~$&$~~m_{\widetilde{u}_{R}}~~$\\	\hline
$	1~	{\rm fb}^{-1}$&$	411	$&$	78	$&$	427	$&$	569	$&$	841	$	\\	\hline
$	5~	{\rm fb}^{-1}$&$	532	$&$	105	$&$	577	$&$	728	$&$	1061	$	\\	\hline
$	8~	{\rm fb}^{-1}$&$	559	$&$	111	$&$	610	$&$	766	$&$	1106	$	\\	\hline
$	10~	{\rm fb}^{-1}$&$	571	$&$	114	  $&$	624	$&$	782	$&$	1125	$	\\	\hline
$	20~ 	{\rm fb}^{-1}$&$	604	$&$	122	$&$	662	$&$	826	$&$	1178	$	\\	\hline
		\end{tabular}
		\label{tab:7TeV_cms}
\end{table}

\begin{table}[ht]
	\centering
	\caption{Reach of the ${\sqrt s}$ = 14 TeV LHC into the No-Scale $\cal{F}$-$SU(5)$ model space utilizing the $S/\sqrt{B+1}\,\ge 5$ condition, using the CMS experiment post-processing cuts and Standard Model background sample of Reference~\cite{PAS-SUS-11-003}. We augment the CMS jet cutting strategy of~\cite{PAS-SUS-11-003} by retaining only those events with greater than or equal to nine jets. All masses are in GeV.}
		\begin{tabular}{|c|c|c|c|c|c|} \hline
${\rm Luminosity}$&$~~M_{1/2}~~$&$~~m_{\widetilde{\chi}^{0}_{1}}~~$&$~~m_{\widetilde{t}_{1}}~~$&$~~m_{\widetilde{g}}~~$&$~~m_{\widetilde{u}_{R}}~~$\\	\hline
$	1~	{\rm fb}^{-1}$&$	695	$&$	143	$&$	768	$&$	945	$&$	1331	$	\\	\hline
$	5~	{\rm fb}^{-1}$&$	774	$&$	162	$&$	860	$&$	1050	$&$	1463	$	\\	\hline
$	8~	{\rm fb}^{-1}$&$	798	$&$	168	$&$	889	$&$	1081	$&$	1505	$	\\	\hline
$	10~	{\rm fb}^{-1}$&$	811	$&$	170	  $&$	903	$&$	1098	$&$	1526	$	\\	\hline
$	20~ 	 {\rm fb}^{-1}$&$	859	$&$	182	$&$	960	$&$	1161	$&$	1610	$	\\	\hline
		\end{tabular}
		\label{tab:14TeV_cms}
\end{table}

\begin{table}[ht]
	\centering
	\caption{Reach of the ${\sqrt s}$ = 7 TeV LHC into the No-Scale $\cal{F}$-$SU(5)$ model space utilizing the $S/\sqrt{B+1}\,\ge 5$ condition, using the ATLAS experiment post-processing cuts and Standard Model background sample of Reference~\cite{Aad:2011qa}. We augment the ATLAS jet cutting strategy of~\cite{Aad:2011qa} by retaining only those events with greater than or equal to seven jets. All masses are in GeV.}
		\begin{tabular}{|c|c|c|c|c|c|} \hline
${\rm Luminosity}$&$~~M_{1/2}~~$&$~~m_{\widetilde{\chi}^{0}_{1}}~~$&$~~m_{\widetilde{t}_{1}}~~$&$~~m_{\widetilde{g}}~~$&$~~m_{\widetilde{u}_{R}}~~$\\	\hline
$	1~	{\rm fb}^{-1}$&$	349	$&$	65	$&$	358	$&$	473	$&$	714	$	\\	\hline
$	5~	{\rm fb}^{-1}$&$	484	$&$	94	$&$	519	$&$	664	$&$	984	$	\\	\hline
$	8~	{\rm fb}^{-1}$&$	522	$&$	102	$&$	565	$&$	715	$&$	1046	$	\\	\hline
$	10~	{\rm fb}^{-1}$&$	537	$&$	106	  $&$	582	$&$	735	$&$	1069	$	\\	\hline
$	20~ 	{\rm fb}^{-1}$&$	574	$&$	115	$&$	627	$&$	785	$&$	1130	$	\\	\hline
		\end{tabular}
		\label{tab:7TeV_atlas}
\end{table}

\begin{table}[ht]
	\centering
	\caption{Reach of the ${\sqrt s}$ = 14 TeV LHC into the No-Scale $\cal{F}$-$SU(5)$ model space utilizing the $S/\sqrt{B+1}\,\ge 5$ condition, using the ATLAS experiment post-processing cuts and Standard Model background sample of Reference~\cite{Aad:2011qa}. We augment the ATLAS jet cutting strategy of~\cite{Aad:2011qa} by retaining only those events with greater than or equal to seven jets. All masses are in GeV.}
		\begin{tabular}{|c|c|c|c|c|c|} \hline
${\rm Luminosity}$&$~~M_{1/2}~~$&$~~m_{\widetilde{\chi}^{0}_{1}}~~$&$~~m_{\widetilde{t}_{1}}~~$&$~~m_{\widetilde{g}}~~$&$~~m_{\widetilde{u}_{R}}~~$\\	\hline
$	1~	{\rm fb}^{-1}$&$	649	$&$	133	$&$	714	$&$	885	$&$	1250	$	\\	\hline
$	5~	{\rm fb}^{-1}$&$	724	$&$	150	$&$	802	$&$	983	$&$	1379	$	\\	\hline
$	8~	{\rm fb}^{-1}$&$	744	$&$	155	$&$	826	$&$	1010	$&$	1414	$	\\	\hline
$	10~	{\rm fb}^{-1}$&$	754	$&$	157	  $&$	836	$&$	1022	$&$	1431	$	\\	\hline
$	20~ 	 {\rm fb}^{-1}$&$	785	$&$	164	$&$	873	$&$	1063	$&$	1484	$	\\	\hline
		\end{tabular}
		\label{tab:14TeV_atlas}
\end{table}

\begin{figure*}[ht]
	\centering
	\includegraphics[width=1.00\textwidth]{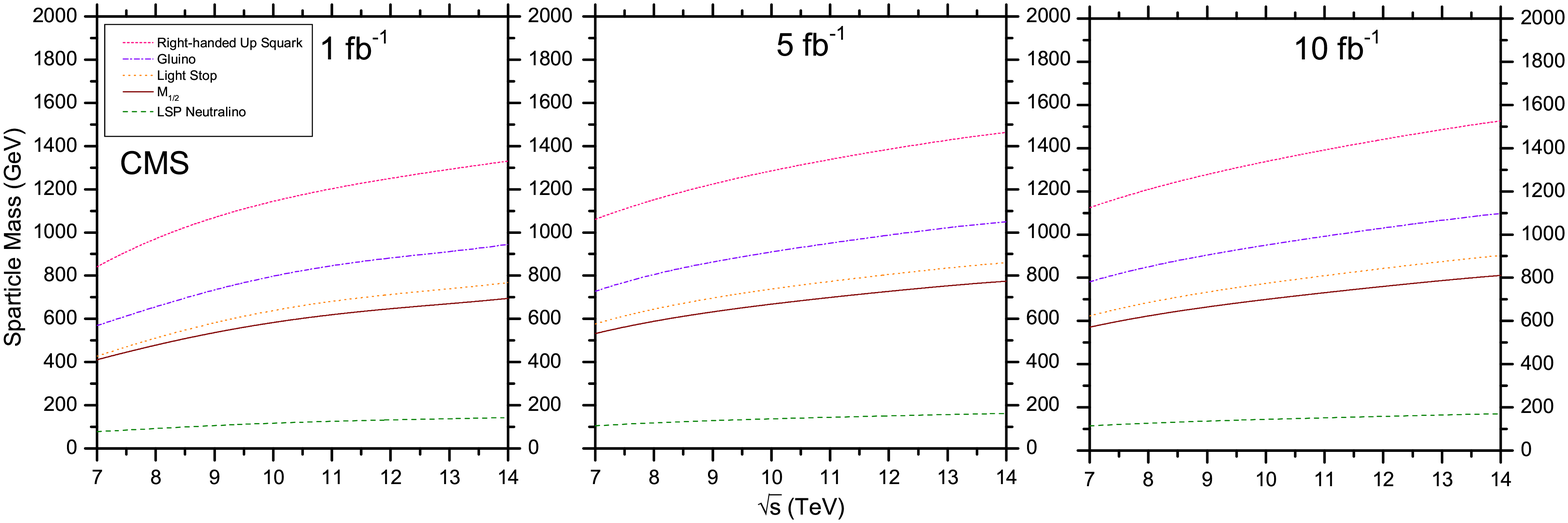}
	\caption{ No-Scale $\cal{F}$-$SU(5)$ sparticle mass discovery threshold as a function of center-of-mass beam energy utilizing the $S/\sqrt{B+1}\,\ge 5$ condition for the three key milestone luminosities of 1 ${\rm fb}^{-1}$, 5 ${\rm fb}^{-1}$, and 10 ${\rm fb}^{-1}$, using the CMS experiment post-processing cuts and Standard Model background sample of Reference~\cite{PAS-SUS-11-003}. We augment the CMS jet cutting strategy of~\cite{PAS-SUS-11-003} by retaining only those events with greater than or equal to nine jets.}
	\label{fig:mass_beamenergy_cms}
\end{figure*}

\begin{figure*}[ht]
	\centering
	\includegraphics[width=1.00\textwidth]{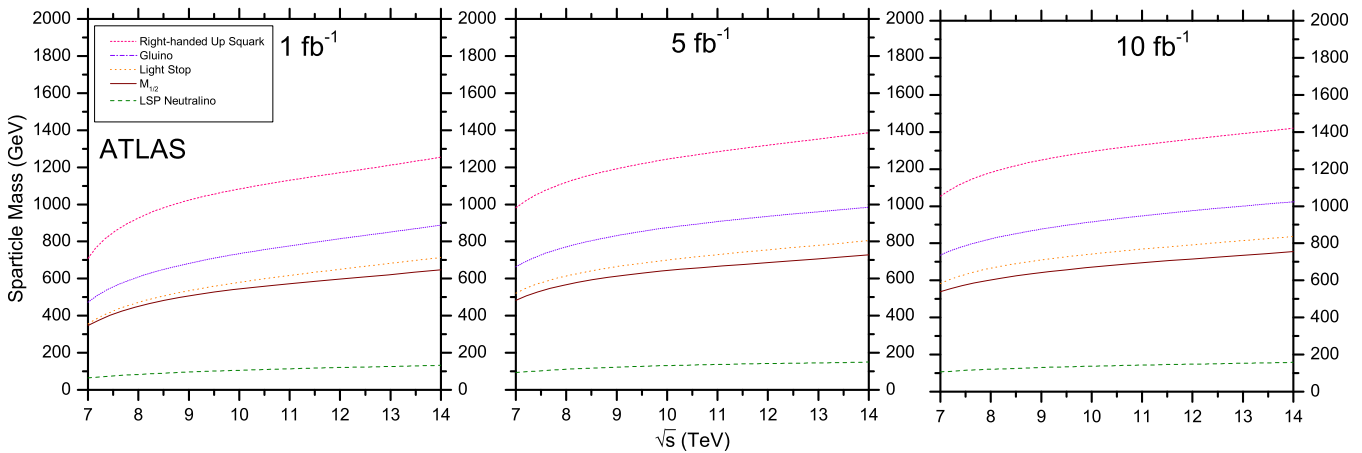}
	\caption{ No-Scale $\cal{F}$-$SU(5)$ sparticle mass discovery threshold as a function of center-of-mass beam energy utilizing the $S/\sqrt{B+1}\,\ge 5$ condition for the three key milestone luminosities of 1 ${\rm fb}^{-1}$, 5 ${\rm fb}^{-1}$, and 10 ${\rm fb}^{-1}$, using the ATLAS experiment post-processing cuts and Standard Model background sample of Reference~\cite{Aad:2011qa}. We augment the ATLAS jet cutting strategy of~\cite{Aad:2011qa} by retaining only those events with greater than or equal to seven jets.}
	\label{fig:mass_beamenergy_atlas}
\end{figure*}

We present an alternative perspective in Figure Sets (\ref{fig:mass_beamenergy_cms}-\ref{fig:mass_beamenergy_atlas}) of the SUSY mass discovery threshold as a function of the center-of-mass beam energy ${\sqrt s}$ for the three key luminosity milestones of 1 ${\rm fb}^{-1}$, 5 ${\rm fb}^{-1}$, and 10 ${\rm fb}^{-1}$. We further submit Tables (\ref{tab:7TeV_cms}-\ref{tab:14TeV_atlas}), reflecting the SUSY mass coverage of the  ${\sqrt s}$ = 7 TeV and ${\sqrt s}$ = 14 TeV LHC for five luminosities of 1 ${\rm fb}^{-1}$, 5 ${\rm fb}^{-1}$, 8 ${\rm fb}^{-1}$,  10 ${\rm fb}^{-1}$,
and 20 ${\rm fb}^{-1}$. In particular, revealed in Table (\ref{tab:7TeV_cms}) at the early LHC run
are the 5 ${\rm fb}^{-1}$ mass targets 
of a 532 GeV gaugino mass $M_{1/2}$, 105 GeV LSP, 577 GeV light stop, 728 GeV gluino, and 1061
GeV right-handed up squark for the CMS experiment, representing a modest infiltration into 
the No-Scale $\cal{F}$-$SU(5)$ model space next year.

It must be emphasized that the entirety of the analysis presented here is highly dependent upon adoption of the ultra-high jet multiplicity cutting methodology described in Refs.~\cite{Maxin:2011hy,Li:2011hr,Li:2011fu,Li:2011av,Li:2011ab}. Here, we have adopted cuts that retain events with only nine jets or more for the CMS experiment jet cutting strategy and only those events with seven jets or more for the ATLAS experiment jet cutting strategy. We have previously demonstrated that more conventional searches which include smaller jet counts than these do not take similar advantage of the strong four-top based decay chain which is distinctively characteristic of the No-Scale $\cal{F}$-$SU(5)$ SUSY mass hierarchy.  As such, all detection thresholds would in this case, be subject to a potentially substantial increase of luminosity.


\section{Conclusions}

The surprising drastic reductions in the viable parameter space experienced by mSUGRA and the CMSSM following only 1 ${\rm fb}^{-1}$ of LHC data compels a study into a more experimentally robust model, such as the No-Scale $\cal{F}$-$SU(5)$ with vector-like particles. We showed that by employing the standard condition $S/\sqrt{B+1}\,\ge 5$ to observe a signal over and above the background, the $\cal{F}$-$SU(5)$ model space had just begun to be probed in the first 1 ${\rm fb}^{-1}$. Consequently, the No-Scale $\cal{F}$-$SU(5)$ remained mostly intact by the early LHC imposed constraints. The $\cal{F}$-$SU(5)$ model did not begin to be pierced until the LHC had attained just under 1 ${\rm fb}^{-1}$, and at the present estimated accumulated luminosity of about 5 ${\rm fb}^{-1}$, modest access to the $\cal{F}$-$SU(5)$ parameter space is currently in progress. While we carry optimism for an early SUSY discovery at a center-of-mass energy of ${\sqrt s}$ = 7 TeV, an absence of a definitive SUSY signal will require the future ${\sqrt s}$ = 14 TeV LHC for full coverage of No-Scale $\cal{F}$-$SU(5)$, along with at least 30 ${\rm fb}^{-1}$ of integrated luminosity. Nevertheless, for an integrated luminosity of 5 ${\rm fb}^{-1}$ at the present ${\sqrt s}$ = 7 TeV LHC, an impressive range of masses will have been reached, including a 532 GeV gaugino mass $M_{1/2}$, a 105 GeV LSP, a 577 GeV light stop, a 728 GeV gluino, and a 1061 GeV right-handed up squark. 

Our recent explorations into all aspects of the No-Scale $\cal{F}$-$SU(5)$ with vector-like particles construction have furnished a remarkable array of deeply motivated theoretical and phenomenological correlations.  We thus suggest that this model makes an excellent concrete choice for the presently undertaken examination of the current and future LHC reach.  The rapid decay of the viable mSUGRA and CMSSM parameter spaces which the early LHC data has brought on moreover suggests the necessity of supplementing such popular and frequently researched SUSY models with phenomenologically and theoretically favorable alternatives, such as No-Scale $\cal{F}$-$SU(5)$.  The ultimate fate of the ongoing search for a genuine supersymmetry signal at the LHC could very well now rest with precisely such outside hopes.


\begin{acknowledgments}
This research was supported in part 
by the DOE grant DE-FG03-95-Er-40917 (TL and DVN),
by the Natural Science Foundation of China 
under grant numbers 10821504 and 11075194 (TL),
by the Mitchell-Heep Chair in High Energy Physics (JAM),
and by the Sam Houston State University
2011 Enhancement Research Grant program (JWW).
We also thank Sam Houston State University
for providing high performance computing resources.
\end{acknowledgments}


\bibliography{bibliography}

\begin{thebibliography}{51}
\expandafter\ifx\csname natexlab\endcsname\relax\def\natexlab#1{#1}\fi
\expandafter\ifx\csname bibnamefont\endcsname\relax
  \def\bibnamefont#1{#1}\fi
\expandafter\ifx\csname bibfnamefont\endcsname\relax
  \def\bibfnamefont#1{#1}\fi
\expandafter\ifx\csname citenamefont\endcsname\relax
  \def\citenamefont#1{#1}\fi
\expandafter\ifx\csname url\endcsname\relax
  \def\url#1{\texttt{#1}}\fi
\expandafter\ifx\csname urlprefix\endcsname\relax\def\urlprefix{URL }\fi
\providecommand{\bibinfo}[2]{#2}
\providecommand{\eprint}[2][]{\url{#2}}

\bibitem[{\citenamefont{Khachatryan et~al.}(2011)}]{Khachatryan:2011tk}
\bibinfo{author}{\bibfnamefont{V.}~\bibnamefont{Khachatryan}}
  \bibnamefont{et~al.} (\bibinfo{collaboration}{CMS Collaboration}),
  {``}\bibinfo{title}{{Search for Supersymmetry in pp Collisions at 7 TeV in
  Events with Jets and Missing Transverse Energy}},{''}
  \bibinfo{journal}{Phys.Lett.} \textbf{\bibinfo{volume}{B698}},
  \bibinfo{pages}{196} (\bibinfo{year}{2011}), \eprint{1101.1628}.

\bibitem[{\citenamefont{Chatrchyan
  et~al.}(2011{\natexlab{a}})}]{Chatrchyan:2011bz}
\bibinfo{author}{\bibfnamefont{S.}~\bibnamefont{Chatrchyan}}
  \bibnamefont{et~al.} (\bibinfo{collaboration}{CMS Collaboration}),
  {``}\bibinfo{title}{{Search for Physics Beyond the Standard Model in
  Opposite-Sign Dilepton Events at $\sqrt{s}$ = 7 TeV}},{''}
  \bibinfo{journal}{JHEP} \textbf{\bibinfo{volume}{1106}}, \bibinfo{pages}{026}
  (\bibinfo{year}{2011}{\natexlab{a}}), \eprint{arXiv:1103.1348}.

\bibitem[{\citenamefont{Chatrchyan
  et~al.}(2011{\natexlab{b}})}]{Chatrchyan:2011bj}
\bibinfo{author}{\bibfnamefont{S.}~\bibnamefont{Chatrchyan}}
  \bibnamefont{et~al.} (\bibinfo{collaboration}{CMS Collaboration}),
  {``}\bibinfo{title}{{Search for Supersymmetry in Events with b Jets and
  Missing Transverse Momentum at the LHC}},{''}
  (\bibinfo{year}{2011}{\natexlab{b}}), \eprint{1106.3272}.

\bibitem[{\citenamefont{Chatrchyan
  et~al.}(2011{\natexlab{c}})}]{Collaboration:2011ida}
\bibinfo{author}{\bibfnamefont{S.}~\bibnamefont{Chatrchyan}}
  \bibnamefont{et~al.} (\bibinfo{collaboration}{CMS Collaboration}),
  {``}\bibinfo{title}{{Search for New Physics with Jets and Missing Transverse
  Momentum in pp collisions at $\sqrt{s}$ = 7 TeV}},{''}
  (\bibinfo{year}{2011}{\natexlab{c}}), \eprint{1106.4503}.

\bibitem[{\citenamefont{Chatrchyan
  et~al.}(2011{\natexlab{d}})}]{Chatrchyan:2011ek}
\bibinfo{author}{\bibfnamefont{S.}~\bibnamefont{Chatrchyan}}
  \bibnamefont{et~al.} (\bibinfo{collaboration}{CMS Collaboration}),
  {``}\bibinfo{title}{{Inclusive search for squarks and gluinos in pp
  collisions at $\sqrt{s}$ = 7 TeV}},{''} (\bibinfo{year}{2011}{\natexlab{d}}),
  \eprint{1107.1279}.

\bibitem[{\citenamefont{CMS}(2011)}]{Collaboration:2011qs}
\bibinfo{author}{\bibnamefont{CMS}}, {``}\bibinfo{title}{{Search for
  supersymmetry in pp collisions at $\sqrt{s}$ = 7 TeV in events with a single
  lepton, jets, and missing transverse momentum}},{''} (\bibinfo{year}{2011}),
  \eprint{1107.1870}.

\bibitem[{PAS(2011)}]{PAS-SUS-11-003}
{``}\bibinfo{title}{{Search for supersymmetry in all-hadronic events with
  $\alpha_{\rm T}$}},{''} (\bibinfo{year}{2011}), \bibinfo{note}{{CMS PAS
  SUS-11-003}}, \urlprefix\url{http://cdsweb.cern.ch/record/1370596}.

\bibitem[{\citenamefont{da~Costa et~al.}(2011{\natexlab{a}})}]{daCosta:2011hh}
\bibinfo{author}{\bibfnamefont{J.~B.~G.} \bibnamefont{da~Costa}}
  \bibnamefont{et~al.} (\bibinfo{collaboration}{Atlas}),
  {``}\bibinfo{title}{{Search for supersymmetry using final states with one
  lepton, jets, and missing transverse momentum with the ATLAS detector in
  $\sqrt{s}$ = 7 TeV pp}},{''} (\bibinfo{year}{2011}{\natexlab{a}}),
  \eprint{1102.2357}.

\bibitem[{\citenamefont{da~Costa et~al.}(2011{\natexlab{b}})}]{daCosta:2011qk}
\bibinfo{author}{\bibfnamefont{J.~B.~G.} \bibnamefont{da~Costa}}
  \bibnamefont{et~al.} (\bibinfo{collaboration}{Atlas}),
  {``}\bibinfo{title}{{Search for squarks and gluinos using final states with
  jets and missing transverse momentum with the ATLAS detector in $\sqrt{s}$ =
  7 TeV proton-proton collisions}},{''} (\bibinfo{year}{2011}{\natexlab{b}}),
  \eprint{1102.5290}.

\bibitem[{\citenamefont{Aad et~al.}(2011{\natexlab{a}})}]{Aad:2011ks}
\bibinfo{author}{\bibfnamefont{G.}~\bibnamefont{Aad}} \bibnamefont{et~al.}
  (\bibinfo{collaboration}{ATLAS Collaboration}), {``}\bibinfo{title}{{Search
  for supersymmetry in pp collisions at $\sqrt{s}$ = 7 TeV in final states with
  missing transverse momentum and b-jets}},{''} \bibinfo{journal}{Phys.\ Lett.\
  B} \textbf{\bibinfo{volume}{B {\bf 701}}}, \bibinfo{pages}{398}
  (\bibinfo{year}{2011}{\natexlab{a}}), \eprint{arXiv:1103.4344}.

\bibitem[{\citenamefont{Aad et~al.}(2011{\natexlab{b}})}]{Aad:2011xm}
\bibinfo{author}{\bibfnamefont{G.}~\bibnamefont{Aad}} \bibnamefont{et~al.}
  (\bibinfo{collaboration}{ATLAS Collaboration}), {``}\bibinfo{title}{{Search
  for supersymmetric particles in events with lepton pairs and large missing
  transverse momentum in $\sqrt{s}$ = 7 TeV proton-proton collisions with the
  ATLAS experiment}},{''} \bibinfo{journal}{Eur.\ Phys.\ J.\ C}
  \textbf{\bibinfo{volume}{C {\bf 71}}}, \bibinfo{pages}{1682}
  (\bibinfo{year}{2011}{\natexlab{b}}), \eprint{arXiv:1103.6214}.

\bibitem[{\citenamefont{ATLAS}(2011)}]{Collaboration:2011jz}
\bibinfo{author}{\bibnamefont{ATLAS}}, {``}\bibinfo{title}{{Measurement of
  dijet production with a veto on additional central jet activity in pp
  collisions at $\sqrt{s}$ = 7 TeV using the ATLAS detector}},{''}
  (\bibinfo{year}{2011}), \eprint{1107.1641}.

\bibitem[{\citenamefont{Aad et~al.}(2011{\natexlab{c}})}]{Aad:2011qa}
\bibinfo{author}{\bibfnamefont{G.}~\bibnamefont{Aad}} \bibnamefont{et~al.}
  (\bibinfo{collaboration}{Atlas}), {``}\bibinfo{title}{{Search for new
  phenomena in final states with large jet multiplicities and missing
  transverse momentum using sqrt(s)=7 TeV pp collisions with the ATLAS
  detector}},{''} (\bibinfo{year}{2011}{\natexlab{c}}), \eprint{1110.2299}.

\bibitem[{\citenamefont{Strumia}(2011)}]{Strumia:2011}
\bibinfo{author}{\bibfnamefont{A.}~\bibnamefont{Strumia}},
  {``}\bibinfo{title}{{Implications of first LHC results}},{''}
  (\bibinfo{year}{2011}), \eprint{1107.1259}.

\bibitem[{\citenamefont{Li et~al.}(2011{\natexlab{a}})\citenamefont{Li, Maxin,
  Nanopoulos, and Walker}}]{Li:2010ws}
\bibinfo{author}{\bibfnamefont{T.}~\bibnamefont{Li}},
  \bibinfo{author}{\bibfnamefont{J.~A.} \bibnamefont{Maxin}},
  \bibinfo{author}{\bibfnamefont{D.~V.} \bibnamefont{Nanopoulos}},
  \bibnamefont{and} \bibinfo{author}{\bibfnamefont{J.~W.}
  \bibnamefont{Walker}}, {``}\bibinfo{title}{{The Golden Point of No-Scale and
  No-Parameter ${\cal F}$-$SU(5)$}},{''} \bibinfo{journal}{Phys. Rev.}
  \textbf{\bibinfo{volume}{D83}}, \bibinfo{pages}{056015}
  (\bibinfo{year}{2011}{\natexlab{a}}), \eprint{1007.5100}.

\bibitem[{\citenamefont{Li et~al.}(2011{\natexlab{b}})\citenamefont{Li, Maxin,
  Nanopoulos, and Walker}}]{Li:2010mi}
\bibinfo{author}{\bibfnamefont{T.}~\bibnamefont{Li}},
  \bibinfo{author}{\bibfnamefont{J.~A.} \bibnamefont{Maxin}},
  \bibinfo{author}{\bibfnamefont{D.~V.} \bibnamefont{Nanopoulos}},
  \bibnamefont{and} \bibinfo{author}{\bibfnamefont{J.~W.}
  \bibnamefont{Walker}}, {``}\bibinfo{title}{{The Golden Strip of Correlated
  Top Quark, Gaugino, and Vectorlike Mass In No-Scale, No-Parameter
  F-SU(5)}},{''} \bibinfo{journal}{Phys. Lett.}
  \textbf{\bibinfo{volume}{B699}}, \bibinfo{pages}{164}
  (\bibinfo{year}{2011}{\natexlab{b}}), \eprint{1009.2981}.

\bibitem[{\citenamefont{Li et~al.}(2011{\natexlab{c}})\citenamefont{Li, Maxin,
  Nanopoulos, and Walker}}]{Li:2010uu}
\bibinfo{author}{\bibfnamefont{T.}~\bibnamefont{Li}},
  \bibinfo{author}{\bibfnamefont{J.~A.} \bibnamefont{Maxin}},
  \bibinfo{author}{\bibfnamefont{D.~V.} \bibnamefont{Nanopoulos}},
  \bibnamefont{and} \bibinfo{author}{\bibfnamefont{J.~W.}
  \bibnamefont{Walker}}, {``}\bibinfo{title}{{Super No-Scale ${\cal
  F}$-$SU(5)$: Resolving the Gauge Hierarchy Problem by Dynamic Determination
  of $M_{1/2}$ and $\tan\beta$}},{''} \bibinfo{journal}{Phys. Lett.}
  \textbf{\bibinfo{volume}{B703}}, \bibinfo{pages}{469}
  (\bibinfo{year}{2011}{\natexlab{c}}), \eprint{1010.4550}.

\bibitem[{\citenamefont{Li et~al.}(2011{\natexlab{d}})\citenamefont{Li, Maxin,
  Nanopoulos, and Walker}}]{Li:2011dw}
\bibinfo{author}{\bibfnamefont{T.}~\bibnamefont{Li}},
  \bibinfo{author}{\bibfnamefont{J.~A.} \bibnamefont{Maxin}},
  \bibinfo{author}{\bibfnamefont{D.~V.} \bibnamefont{Nanopoulos}},
  \bibnamefont{and} \bibinfo{author}{\bibfnamefont{J.~W.}
  \bibnamefont{Walker}}, {``}\bibinfo{title}{{Blueprints of the No-Scale
  Multiverse at the LHC}},{''} \bibinfo{journal}{Phys. Rev.}
  \textbf{\bibinfo{volume}{D84}}, \bibinfo{pages}{056016}
  (\bibinfo{year}{2011}{\natexlab{d}}), \eprint{1101.2197}.

\bibitem[{\citenamefont{Li et~al.}(2011{\natexlab{e}})\citenamefont{Li, Maxin,
  Nanopoulos, and Walker}}]{Li:2011hr}
\bibinfo{author}{\bibfnamefont{T.}~\bibnamefont{Li}},
  \bibinfo{author}{\bibfnamefont{J.~A.} \bibnamefont{Maxin}},
  \bibinfo{author}{\bibfnamefont{D.~V.} \bibnamefont{Nanopoulos}},
  \bibnamefont{and} \bibinfo{author}{\bibfnamefont{J.~W.}
  \bibnamefont{Walker}}, {``}\bibinfo{title}{{Ultra High Jet Signals from
  Stringy No-Scale Supergravity}},{''} (\bibinfo{year}{2011}{\natexlab{e}}),
  \eprint{1103.2362}.

\bibitem[{\citenamefont{Li et~al.}(2011{\natexlab{f}})\citenamefont{Li, Maxin,
  Nanopoulos, and Walker}}]{Maxin:2011hy}
\bibinfo{author}{\bibfnamefont{T.}~\bibnamefont{Li}},
  \bibinfo{author}{\bibfnamefont{J.~A.} \bibnamefont{Maxin}},
  \bibinfo{author}{\bibfnamefont{D.~V.} \bibnamefont{Nanopoulos}},
  \bibnamefont{and} \bibinfo{author}{\bibfnamefont{J.~W.}
  \bibnamefont{Walker}}, {``}\bibinfo{title}{{The Ultrahigh jet multiplicity
  signal of stringy no-scale $\cal{F}$-$SU(5)$ at the $\sqrt{s}= 7$ TeV
  LHC}},{''} \bibinfo{journal}{Phys.Rev.} \textbf{\bibinfo{volume}{D84}},
  \bibinfo{pages}{076003} (\bibinfo{year}{2011}{\natexlab{f}}),
  \eprint{1103.4160}.

\bibitem[{\citenamefont{Li et~al.}(2011{\natexlab{g}})\citenamefont{Li, Maxin,
  Nanopoulos, and Walker}}]{Li:2011xu}
\bibinfo{author}{\bibfnamefont{T.}~\bibnamefont{Li}},
  \bibinfo{author}{\bibfnamefont{J.~A.} \bibnamefont{Maxin}},
  \bibinfo{author}{\bibfnamefont{D.~V.} \bibnamefont{Nanopoulos}},
  \bibnamefont{and} \bibinfo{author}{\bibfnamefont{J.~W.}
  \bibnamefont{Walker}}, {``}\bibinfo{title}{{The Unification of Dynamical
  Determination and Bare Minimal Phenomenological Constraints in No-Scale
  \cal{F}- SU(5)}},{''} (\bibinfo{year}{2011}{\natexlab{g}}),
  \eprint{1105.3988}.

\bibitem[{\citenamefont{Li et~al.}(2011{\natexlab{h}})\citenamefont{Li, Maxin,
  Nanopoulos, and Walker}}]{Li:2011in}
\bibinfo{author}{\bibfnamefont{T.}~\bibnamefont{Li}},
  \bibinfo{author}{\bibfnamefont{J.~A.} \bibnamefont{Maxin}},
  \bibinfo{author}{\bibfnamefont{D.~V.} \bibnamefont{Nanopoulos}},
  \bibnamefont{and} \bibinfo{author}{\bibfnamefont{J.~W.}
  \bibnamefont{Walker}}, {``}\bibinfo{title}{{The Race for Supersymmetric Dark
  Matter at XENON100 and the LHC: Stringy Correlations from No-Scale
  \cal{F}-SU(5)}},{''} (\bibinfo{year}{2011}{\natexlab{h}}),
  \eprint{1106.1165}.

\bibitem[{\citenamefont{Li et~al.}(2011{\natexlab{i}})\citenamefont{Li, Maxin,
  Nanopoulos, and Walker}}]{Li:2011gh}
\bibinfo{author}{\bibfnamefont{T.}~\bibnamefont{Li}},
  \bibinfo{author}{\bibfnamefont{J.~A.} \bibnamefont{Maxin}},
  \bibinfo{author}{\bibfnamefont{D.~V.} \bibnamefont{Nanopoulos}},
  \bibnamefont{and} \bibinfo{author}{\bibfnamefont{J.~W.}
  \bibnamefont{Walker}}, {``}\bibinfo{title}{{A Two-Tiered Correlation of Dark
  Matter with Missing Transverse Energy: Reconstructing the Lightest
  Supersymmetric Particle Mass at the LHC}},{''}
  (\bibinfo{year}{2011}{\natexlab{i}}), \eprint{1107.2375}.

\bibitem[{\citenamefont{Li et~al.}(2011{\natexlab{j}})\citenamefont{Li, Maxin,
  Nanopoulos, and Walker}}]{Li:2011fu}
\bibinfo{author}{\bibfnamefont{T.}~\bibnamefont{Li}},
  \bibinfo{author}{\bibfnamefont{J.~A.} \bibnamefont{Maxin}},
  \bibinfo{author}{\bibfnamefont{D.~V.} \bibnamefont{Nanopoulos}},
  \bibnamefont{and} \bibinfo{author}{\bibfnamefont{J.~W.}
  \bibnamefont{Walker}}, {``}\bibinfo{title}{{Has SUSY Gone Undetected in 9-jet
  Events? A Ten-Fold Enhancement in the LHC Signal Efficiency}},{''}
  (\bibinfo{year}{2011}{\natexlab{j}}), \eprint{1108.5169}.

\bibitem[{\citenamefont{Li et~al.}(2012)\citenamefont{Li, Maxin, Nanopoulos,
  and Walker}}]{Li:2011xg}
\bibinfo{author}{\bibfnamefont{T.}~\bibnamefont{Li}},
  \bibinfo{author}{\bibfnamefont{J.~A.} \bibnamefont{Maxin}},
  \bibinfo{author}{\bibfnamefont{D.~V.} \bibnamefont{Nanopoulos}},
  \bibnamefont{and} \bibinfo{author}{\bibfnamefont{J.~W.}
  \bibnamefont{Walker}}, {``}\bibinfo{title}{{Natural Predictions for the Higgs
  Boson Mass and Supersymmetric Contributions to Rare Processes}},{''}
  \bibinfo{journal}{Phys.Lett.} \textbf{\bibinfo{volume}{B708}},
  \bibinfo{pages}{93} (\bibinfo{year}{2012}), \eprint{1109.2110}.

\bibitem[{\citenamefont{Li et~al.}(2011{\natexlab{k}})\citenamefont{Li, Maxin,
  Nanopoulos, and Walker}}]{Li:2011ex}
\bibinfo{author}{\bibfnamefont{T.}~\bibnamefont{Li}},
  \bibinfo{author}{\bibfnamefont{J.~A.} \bibnamefont{Maxin}},
  \bibinfo{author}{\bibfnamefont{D.~V.} \bibnamefont{Nanopoulos}},
  \bibnamefont{and} \bibinfo{author}{\bibfnamefont{J.~W.}
  \bibnamefont{Walker}}, {``}\bibinfo{title}{{The F-Landscape: Dynamically
  Determining the Multiverse}},{''} (\bibinfo{year}{2011}{\natexlab{k}}),
  \eprint{1111.0236}.

\bibitem[{\citenamefont{Li et~al.}(2011{\natexlab{l}})\citenamefont{Li, Maxin,
  Nanopoulos, and Walker}}]{Li:2011av}
\bibinfo{author}{\bibfnamefont{T.}~\bibnamefont{Li}},
  \bibinfo{author}{\bibfnamefont{J.~A.} \bibnamefont{Maxin}},
  \bibinfo{author}{\bibfnamefont{D.~V.} \bibnamefont{Nanopoulos}},
  \bibnamefont{and} \bibinfo{author}{\bibfnamefont{J.~W.}
  \bibnamefont{Walker}}, {``}\bibinfo{title}{{Profumo di SUSY: Suggestive
  Correlations in the ATLAS and CMS High Jet Multiplicity Data}},{''}
  (\bibinfo{year}{2011}{\natexlab{l}}), \eprint{1111.4204}.

\bibitem[{\citenamefont{Li et~al.}(2011{\natexlab{m}})\citenamefont{Li, Maxin,
  Nanopoulos, and Walker}}]{Li:2011ab}
\bibinfo{author}{\bibfnamefont{T.}~\bibnamefont{Li}},
  \bibinfo{author}{\bibfnamefont{J.~A.} \bibnamefont{Maxin}},
  \bibinfo{author}{\bibfnamefont{D.~V.} \bibnamefont{Nanopoulos}},
  \bibnamefont{and} \bibinfo{author}{\bibfnamefont{J.~W.}
  \bibnamefont{Walker}}, {``}\bibinfo{title}{{A Higgs Mass Shift to 125 GeV and
  A Multi-Jet Supersymmetry Signal: Miracle of the Flippons at the $\sqrt{s} =
  7$~TeV LHC}},{''} (\bibinfo{year}{2011}{\natexlab{m}}), \eprint{1112.3024}.

\bibitem[{\citenamefont{Cremmer et~al.}(1983)\citenamefont{Cremmer, Ferrara,
  Kounnas, and Nanopoulos}}]{Cremmer:1983bf}
\bibinfo{author}{\bibfnamefont{E.}~\bibnamefont{Cremmer}},
  \bibinfo{author}{\bibfnamefont{S.}~\bibnamefont{Ferrara}},
  \bibinfo{author}{\bibfnamefont{C.}~\bibnamefont{Kounnas}}, \bibnamefont{and}
  \bibinfo{author}{\bibfnamefont{D.~V.} \bibnamefont{Nanopoulos}},
  {``}\bibinfo{title}{{Naturally Vanishing Cosmological Constant in $N=1$
  Supergravity}},{''} \bibinfo{journal}{Phys. Lett.}
  \textbf{\bibinfo{volume}{B133}}, \bibinfo{pages}{61} (\bibinfo{year}{1983}).

\bibitem[{\citenamefont{Ellis et~al.}(1984{\natexlab{a}})\citenamefont{Ellis,
  Lahanas, Nanopoulos, and Tamvakis}}]{Ellis:1983sf}
\bibinfo{author}{\bibfnamefont{J.~R.} \bibnamefont{Ellis}},
  \bibinfo{author}{\bibfnamefont{A.~B.} \bibnamefont{Lahanas}},
  \bibinfo{author}{\bibfnamefont{D.~V.} \bibnamefont{Nanopoulos}},
  \bibnamefont{and} \bibinfo{author}{\bibfnamefont{K.}~\bibnamefont{Tamvakis}},
  {``}\bibinfo{title}{{No-Scale Supersymmetric Standard Model}},{''}
  \bibinfo{journal}{Phys. Lett.} \textbf{\bibinfo{volume}{B134}},
  \bibinfo{pages}{429} (\bibinfo{year}{1984}{\natexlab{a}}).

\bibitem[{\citenamefont{Ellis et~al.}(1984{\natexlab{b}})\citenamefont{Ellis,
  Kounnas, and Nanopoulos}}]{Ellis:1983ei}
\bibinfo{author}{\bibfnamefont{J.~R.} \bibnamefont{Ellis}},
  \bibinfo{author}{\bibfnamefont{C.}~\bibnamefont{Kounnas}}, \bibnamefont{and}
  \bibinfo{author}{\bibfnamefont{D.~V.} \bibnamefont{Nanopoulos}},
  {``}\bibinfo{title}{{Phenomenological $SU(1,1)$ Supergravity}},{''}
  \bibinfo{journal}{Nucl. Phys.} \textbf{\bibinfo{volume}{B241}},
  \bibinfo{pages}{406} (\bibinfo{year}{1984}{\natexlab{b}}).

\bibitem[{\citenamefont{Ellis et~al.}(1984{\natexlab{c}})\citenamefont{Ellis,
  Kounnas, and Nanopoulos}}]{Ellis:1984bm}
\bibinfo{author}{\bibfnamefont{J.~R.} \bibnamefont{Ellis}},
  \bibinfo{author}{\bibfnamefont{C.}~\bibnamefont{Kounnas}}, \bibnamefont{and}
  \bibinfo{author}{\bibfnamefont{D.~V.} \bibnamefont{Nanopoulos}},
  {``}\bibinfo{title}{{No Scale Supersymmetric Guts}},{''}
  \bibinfo{journal}{Nucl. Phys.} \textbf{\bibinfo{volume}{B247}},
  \bibinfo{pages}{373} (\bibinfo{year}{1984}{\natexlab{c}}).

\bibitem[{\citenamefont{Lahanas and Nanopoulos}(1987)}]{Lahanas:1986uc}
\bibinfo{author}{\bibfnamefont{A.~B.} \bibnamefont{Lahanas}} \bibnamefont{and}
  \bibinfo{author}{\bibfnamefont{D.~V.} \bibnamefont{Nanopoulos}},
  {``}\bibinfo{title}{{The Road to No Scale Supergravity}},{''}
  \bibinfo{journal}{Phys. Rept.} \textbf{\bibinfo{volume}{145}},
  \bibinfo{pages}{1} (\bibinfo{year}{1987}).

\bibitem[{\citenamefont{Barr}(1982)}]{Barr:1981qv}
\bibinfo{author}{\bibfnamefont{S.~M.} \bibnamefont{Barr}},
  {``}\bibinfo{title}{{A New Symmetry Breaking Pattern for $SO(10)$ and Proton
  Decay}},{''} \bibinfo{journal}{Phys. Lett.} \textbf{\bibinfo{volume}{B112}},
  \bibinfo{pages}{219} (\bibinfo{year}{1982}).

\bibitem[{\citenamefont{Derendinger et~al.}(1984)\citenamefont{Derendinger,
  Kim, and Nanopoulos}}]{Derendinger:1983aj}
\bibinfo{author}{\bibfnamefont{J.~P.} \bibnamefont{Derendinger}},
  \bibinfo{author}{\bibfnamefont{J.~E.} \bibnamefont{Kim}}, \bibnamefont{and}
  \bibinfo{author}{\bibfnamefont{D.~V.} \bibnamefont{Nanopoulos}},
  {``}\bibinfo{title}{{Anti-$SU(5)$}},{''} \bibinfo{journal}{Phys. Lett.}
  \textbf{\bibinfo{volume}{B139}}, \bibinfo{pages}{170} (\bibinfo{year}{1984}).

\bibitem[{\citenamefont{Antoniadis et~al.}(1987)\citenamefont{Antoniadis,
  Ellis, Hagelin, and Nanopoulos}}]{Antoniadis:1987dx}
\bibinfo{author}{\bibfnamefont{I.}~\bibnamefont{Antoniadis}},
  \bibinfo{author}{\bibfnamefont{J.~R.} \bibnamefont{Ellis}},
  \bibinfo{author}{\bibfnamefont{J.~S.} \bibnamefont{Hagelin}},
  \bibnamefont{and} \bibinfo{author}{\bibfnamefont{D.~V.}
  \bibnamefont{Nanopoulos}}, {``}\bibinfo{title}{{Supersymmetric Flipped
  $SU(5)$ Revitalized}},{''} \bibinfo{journal}{Phys. Lett.}
  \textbf{\bibinfo{volume}{B194}}, \bibinfo{pages}{231} (\bibinfo{year}{1987}).

\bibitem[{\citenamefont{Jiang et~al.}(2007)\citenamefont{Jiang, Li, and
  Nanopoulos}}]{Jiang:2006hf}
\bibinfo{author}{\bibfnamefont{J.}~\bibnamefont{Jiang}},
  \bibinfo{author}{\bibfnamefont{T.}~\bibnamefont{Li}}, \bibnamefont{and}
  \bibinfo{author}{\bibfnamefont{D.~V.} \bibnamefont{Nanopoulos}},
  {``}\bibinfo{title}{{Testable Flipped $SU(5) \times U(1)_X$ Models}},{''}
  \bibinfo{journal}{Nucl. Phys.} \textbf{\bibinfo{volume}{B772}},
  \bibinfo{pages}{49} (\bibinfo{year}{2007}), \eprint{hep-ph/0610054}.

\bibitem[{\citenamefont{Jiang et~al.}(2009)\citenamefont{Jiang, Li, Nanopoulos,
  and Xie}}]{Jiang:2009zza}
\bibinfo{author}{\bibfnamefont{J.}~\bibnamefont{Jiang}},
  \bibinfo{author}{\bibfnamefont{T.}~\bibnamefont{Li}},
  \bibinfo{author}{\bibfnamefont{D.~V.} \bibnamefont{Nanopoulos}},
  \bibnamefont{and} \bibinfo{author}{\bibfnamefont{D.}~\bibnamefont{Xie}},
  {``}\bibinfo{title}{{F-$SU(5)$}},{''} \bibinfo{journal}{Phys. Lett.}
  \textbf{\bibinfo{volume}{B677}}, \bibinfo{pages}{322} (\bibinfo{year}{2009}).

\bibitem[{\citenamefont{Jiang et~al.}(2010)\citenamefont{Jiang, Li, Nanopoulos,
  and Xie}}]{Jiang:2009za}
\bibinfo{author}{\bibfnamefont{J.}~\bibnamefont{Jiang}},
  \bibinfo{author}{\bibfnamefont{T.}~\bibnamefont{Li}},
  \bibinfo{author}{\bibfnamefont{D.~V.} \bibnamefont{Nanopoulos}},
  \bibnamefont{and} \bibinfo{author}{\bibfnamefont{D.}~\bibnamefont{Xie}},
  {``}\bibinfo{title}{{Flipped $SU(5) \times U(1)_X$ Models from
  F-Theory}},{''} \bibinfo{journal}{Nucl. Phys.}
  \textbf{\bibinfo{volume}{B830}}, \bibinfo{pages}{195} (\bibinfo{year}{2010}),
  \eprint{0905.3394}.

\bibitem[{\citenamefont{Li et~al.}(2011{\natexlab{n}})\citenamefont{Li,
  Nanopoulos, and Walker}}]{Li:2010dp}
\bibinfo{author}{\bibfnamefont{T.}~\bibnamefont{Li}},
  \bibinfo{author}{\bibfnamefont{D.~V.} \bibnamefont{Nanopoulos}},
  \bibnamefont{and} \bibinfo{author}{\bibfnamefont{J.~W.}
  \bibnamefont{Walker}}, {``}\bibinfo{title}{{Elements of F-ast Proton
  Decay}},{''} \bibinfo{journal}{Nucl. Phys.} \textbf{\bibinfo{volume}{B846}},
  \bibinfo{pages}{43} (\bibinfo{year}{2011}{\natexlab{n}}), \eprint{1003.2570}.

\bibitem[{\citenamefont{Li et~al.}(2011{\natexlab{o}})\citenamefont{Li, Maxin,
  Nanopoulos, and Walker}}]{Li:2010rz}
\bibinfo{author}{\bibfnamefont{T.}~\bibnamefont{Li}},
  \bibinfo{author}{\bibfnamefont{J.~A.} \bibnamefont{Maxin}},
  \bibinfo{author}{\bibfnamefont{D.~V.} \bibnamefont{Nanopoulos}},
  \bibnamefont{and} \bibinfo{author}{\bibfnamefont{J.~W.}
  \bibnamefont{Walker}}, {``}\bibinfo{title}{{Dark Matter, Proton Decay and
  Other Phenomenological Constraints in ${\cal F}$-SU(5)}},{''}
  \bibinfo{journal}{Nucl.Phys.} \textbf{\bibinfo{volume}{B848}},
  \bibinfo{pages}{314} (\bibinfo{year}{2011}{\natexlab{o}}),
  \eprint{1003.4186}.

\bibitem[{\citenamefont{Li et~al.}(2010)\citenamefont{Li, Nanopoulos, and
  Walker}}]{Li:2009fq}
\bibinfo{author}{\bibfnamefont{T.}~\bibnamefont{Li}},
  \bibinfo{author}{\bibfnamefont{D.~V.} \bibnamefont{Nanopoulos}},
  \bibnamefont{and} \bibinfo{author}{\bibfnamefont{J.~W.}
  \bibnamefont{Walker}}, {``}\bibinfo{title}{{Fast Proton Decay}},{''}
  \bibinfo{journal}{Phys. Lett.} \textbf{\bibinfo{volume}{B693}},
  \bibinfo{pages}{580} (\bibinfo{year}{2010}), \eprint{0910.0860}.

\bibitem[{\citenamefont{Allanach et~al.}(2002)}]{Allanach:2002nj}
\bibinfo{author}{\bibfnamefont{B.~C.} \bibnamefont{Allanach}}
  \bibnamefont{et~al.}, {``}\bibinfo{title}{{The Snowmass points and slopes:
  Benchmarks for SUSY searches}},{''} \bibinfo{journal}{Eur. Phys. J.}
  \textbf{\bibinfo{volume}{C25}}, \bibinfo{pages}{113} (\bibinfo{year}{2002}),
  \eprint{hep-ph/0202233}.

\bibitem[{\citenamefont{Stelzer and Long}(1994)}]{Stelzer:1994ta}
\bibinfo{author}{\bibfnamefont{T.}~\bibnamefont{Stelzer}} \bibnamefont{and}
  \bibinfo{author}{\bibfnamefont{W.~F.} \bibnamefont{Long}},
  {``}\bibinfo{title}{{Automatic generation of tree level helicity
  amplitudes}},{''} \bibinfo{journal}{Comput. Phys. Commun.}
  \textbf{\bibinfo{volume}{81}}, \bibinfo{pages}{357} (\bibinfo{year}{1994}),
  \eprint{hep-ph/9401258}.

\bibitem[{\citenamefont{Alwall et~al.}(2011)}]{MGME}
\bibinfo{author}{\bibfnamefont{J.}~\bibnamefont{Alwall}} \bibnamefont{et~al.},
  {``}\bibinfo{title}{MadGraph/MadEvent Collider Event Simulation Suite},{''}
  (\bibinfo{year}{2011}), \urlprefix\url{http://madgraph.hep.uiuc.edu/}.

\bibitem[{\citenamefont{Alwall et~al.}(2007)}]{Alwall:2007st}
\bibinfo{author}{\bibfnamefont{J.}~\bibnamefont{Alwall}} \bibnamefont{et~al.},
  {``}\bibinfo{title}{{MadGraph/MadEvent v4: The New Web Generation}},{''}
  \bibinfo{journal}{JHEP} \textbf{\bibinfo{volume}{09}}, \bibinfo{pages}{028}
  (\bibinfo{year}{2007}), \eprint{0706.2334}.

\bibitem[{\citenamefont{Sjostrand et~al.}(2006)\citenamefont{Sjostrand, Mrenna,
  and Skands}}]{Sjostrand:2006za}
\bibinfo{author}{\bibfnamefont{T.}~\bibnamefont{Sjostrand}},
  \bibinfo{author}{\bibfnamefont{S.}~\bibnamefont{Mrenna}}, \bibnamefont{and}
  \bibinfo{author}{\bibfnamefont{P.~Z.} \bibnamefont{Skands}},
  {``}\bibinfo{title}{{PYTHIA 6.4 Physics and Manual}},{''}
  \bibinfo{journal}{JHEP} \textbf{\bibinfo{volume}{05}}, \bibinfo{pages}{026}
  (\bibinfo{year}{2006}), \eprint{hep-ph/0603175}.

\bibitem[{\citenamefont{Conway et~al.}(2009)}]{PGS4}
\bibinfo{author}{\bibfnamefont{J.}~\bibnamefont{Conway}} \bibnamefont{et~al.},
  {``}\bibinfo{title}{PGS4: Pretty Good (Detector) Simulation},{''}
  (\bibinfo{year}{2009}),
  \urlprefix\url{http://www.physics.ucdavis.edu/~conway/research/}.

\bibitem[{\citenamefont{Li et~al.}(2011{\natexlab{p}})\citenamefont{Li, Maxin,
  Nanopoulos, and Walker}}]{cutlhco}
\bibinfo{author}{\bibfnamefont{T.}~\bibnamefont{Li}},
  \bibinfo{author}{\bibfnamefont{J.~A.} \bibnamefont{Maxin}},
  \bibinfo{author}{\bibfnamefont{D.~V.} \bibnamefont{Nanopoulos}},
  \bibnamefont{and} \bibinfo{author}{\bibfnamefont{J.~W.}
  \bibnamefont{Walker}}, {``}\bibinfo{title}{CutLHCO: A Tool For Detector
  Selection Cuts},{''} (\bibinfo{year}{2011}{\natexlab{p}}),
  \urlprefix\url{http://www.joelwalker.net/code/cut_lhco.tar.gz}.

\bibitem[{\citenamefont{Djouadi et~al.}(2007)\citenamefont{Djouadi, Kneur, and
  Moultaka}}]{Djouadi:2002ze}
\bibinfo{author}{\bibfnamefont{A.}~\bibnamefont{Djouadi}},
  \bibinfo{author}{\bibfnamefont{J.-L.} \bibnamefont{Kneur}}, \bibnamefont{and}
  \bibinfo{author}{\bibfnamefont{G.}~\bibnamefont{Moultaka}},
  {``}\bibinfo{title}{{SuSpect: A Fortran code for the supersymmetric and Higgs
  particle spectrum in the MSSM}},{''} \bibinfo{journal}{Comput. Phys. Commun.}
  \textbf{\bibinfo{volume}{176}}, \bibinfo{pages}{426} (\bibinfo{year}{2007}),
  \eprint{hep-ph/0211331}.

\bibitem[{\citenamefont{Belanger et~al.}(2009)\citenamefont{Belanger, Boudjema,
  Pukhov, and Semenov}}]{Belanger:2008sj}
\bibinfo{author}{\bibfnamefont{G.}~\bibnamefont{Belanger}},
  \bibinfo{author}{\bibfnamefont{F.}~\bibnamefont{Boudjema}},
  \bibinfo{author}{\bibfnamefont{A.}~\bibnamefont{Pukhov}}, \bibnamefont{and}
  \bibinfo{author}{\bibfnamefont{A.}~\bibnamefont{Semenov}},
  {``}\bibinfo{title}{{Dark matter direct detection rate in a generic model
  with micrOMEGAs2.1}},{''} \bibinfo{journal}{Comput. Phys. Commun.}
  \textbf{\bibinfo{volume}{180}}, \bibinfo{pages}{747} (\bibinfo{year}{2009}),
  \eprint{0803.2360}.

\end{thebibliography}

\end{document}